# Systematically reviewing the layered architectural pattern principles and their use to reconstruct software architectures

ALVINE B. BELLE*

Lassonde School of Engineering, York University (Canada)

GHIZLANE EL BOUSSAIDI

Department of Software and IT Engineering, École de Technologie Supérieure (Canada)

TIMOTHY C. LETHBRIDGE

School of Electrical Engineering and Computer Science, University of Ottawa (Canada)

SEGLA KPODJEDO

Department of Software and IT Engineering, École de Technologie Supérieure (Canada)

HAFEDH MILI

LATECE Laboratory, Université du Québec à Montréal, Montréal, Canada

ANDRÉS PAZ

Department of Software and IT Engineering, École de Technologie Supérieure (Canada)

**Abstract**—Architectural reconstruction is a reverse engineering activity aiming at recovering the missing decisions on a system. It can help identify the components, within a legacy software application, according to the application's architectural pattern. It is useful to identify architectural technical debt. We are interested in identifying layers within a layered application since the layered pattern is one of the most used patterns to structure large systems. Earlier component reconstruction work focusing on that pattern relied on generic component identification criteria, such as cohesion and coupling. Recent work has identified architectural-pattern specific criteria to identify components within that pattern. However, the architectural-pattern specific criteria that the layered pattern embodies are loosely defined. In this paper, we present a first systematic literature review (SLR) of the literature aiming at inventorying such criteria for layers within legacy applications and grouping them under four principles that embody the fundamental design principles underlying the architectural pattern. We identify six such criteria in the form of design rules. We also perform a second systematic literature review to synthesize the literature on software architecture reconstruction in the light of these criteria. We report those principles, the rules they encompass, their representation, and their usage in software architecture reconstruction.

## 1. INTRODUCTION

The architecture of a software application is embodied in its components, their behavior, and their interactions [1, 150]. Similar architectures may be defined in terms of *architectural patterns*, which are described in terms of component *types, connectors*, and their relationships [2]. Given a legacy application, it is difficult to identify the components of the application. In fact, the *component types* may not have recognizable, language-supported boundaries within the legacy code. Moreover, documentation, when such exists, may not be accurate either from day one [3, 5, 6] or due to repeated undisciplined changes to the software [3, 4]. Thus, we normally need to *reconstruct* the architecture of a system by analyzing its source code [4, 7].

* Corresponding author. Email = alvine.boayebelle@mail.mcgill.ca

The literature refers to software architecture reconstruction using different terms: reverse architecting, or architecture extraction, mining, recovery, or discovery. In this context, *discovery* generally designates a top-down process while *recovery* refers to a bottom-up process [4]. Architecture reconstruction is akin to recovery in that it generally starts from source code and progressively constructs a more abstract representation of the system.

The reconstructed architecture serves as a medium to carry out diverse analyses such as patterns conformance, dependency analysis, or quality attribute analysis [4]. Such analyses give useful information easing the design decision-making process throughout the various stages of the software maintenance and evolution. They can be a means to identify technical debt and, more specifically, architectural technical debt (ATD) since they can help detect software architecture degradation. That erosion of the system's design makes it brittle, immobile, difficult and expensive to maintain, upgrade, and understand. Besides, it may result in a concrete architecture that clashes with the system's conceptual architecture [38].

The idea behind rule-based architecture reconstruction is to uncover the architecture using the same design rules that were used for its original design. In fact, architectural design typically follows several rules to achieve desirable properties. Adherence to these rules imparts certain properties of the components of the architecture. It hence makes sense to uncover components, or component boundaries, by looking for manifestations of the properties or qualities imparted by these rules.

To the extent that component identification is an important aspect of architecture reconstruction, early work on architecture reconstruction relied on general principles used for *modular design* (e.g., [8-11, 14]), identifying components by looking for clusters of functionality that, considered together, exhibit such principles. These general principles include *separation of concerns*, *information hiding*, *low coupling*, *high cohesion*, *protected variations*, *encapsulation*, *abstraction*, the *open-closed* principle and the *acyclic dependencies* principle[1] (e.g., [12, 13, 32, 35, 36, 101, 125]). However, such criteria not only lack *precision*—in the sense of *not* distinguishing between different *component types*—but may not even be correct [8]. For example, *utility components* that implement cross-cutting infrastructure functionality within an application (e.g., logging) will exhibit neither high cohesion nor low coupling. Furthermore, even though software architecture reconstruction approaches are inaccurate in general [81], those approaches that focus on applying the above general principles are the ones that seem to lack accuracy the most, as demonstrated by the empirical study of Garcia et al. [69]. In particular, the analyses that these authors carried out revealed that the poor accuracy of such approaches suggests that the criteria they used while reconstructing software architectures do not reflect the way engineers assigned functionality to components. Hence, the architectures reconstructed using these general principles might not properly reflect the architectural patterns used to create them. This means that these architectures might not convey the architectural significance or the splitting mechanisms embodied by the architectural pattern used to create them [17]. Accordingly, more recent work relied on specific rules of architectural patterns to reconstruct the architectures of systems that are known to be layered systems (e.g., [8, 15, 16, 33]).

In this paper, we choose to focus on the layered pattern, which is one of the most used architectural patterns and that is widely adopted by the developer community to structure large industrial systems [15]. The literature usually refers to that pattern as the ''layered architectural style'', the ''layered view'' or even the ''layering'' [15]. When it comes to the layered pattern, literature implicitly or explicitly leverages several rules, including: the *closed layering* rule (e.g., [15 and 28]), the *open layering* rule (e.g., [16, 29]), the *skip-call (ban)* rule (e.g., [15, 89]), the *back-call (ban) rule* (e.g., [15, 89]), and the *acyclic dependency* rule (e.g., [19, 89]). For instance, most of the software architecture reconstruction approaches use heuristics to ignore parts of the analyzed system (mostly dependencies) which violate the layered pattern and build the layered architecture using the resulting acyclic representation of the system.

A survey conducted by Savolainen and Myllarniemi [70] in 2009 showed that the usage of the layered architectural pattern differs greatly depending on the context. This drove them to compare how the notion of software architecture layers was perceived in research literature as well as in industrial practice. Their survey acknowledged that the layered pattern is widely used. The survey results showed that several papers discussed layers in various contexts, but the research done on layers' notions and concepts is surprisingly scarce. Besides, most literature describing the layered pattern reuses the same seminal work (e.g., [2, 13, 21, 71, 72]) and the pattern descriptions and principles they propose are sometimes vague and lack uniformity. More recently, Pruijt et al. [105] argue that the layered pattern lacks a uniform classification: distinct authors use varying and sometimes contradictory terms for layers, logic types (e.g., "application logic", "business logic") as well as responsibility types. Several papers have been published in the meantime. However, their main issues are that the rules they convey usually lack uniformity and are loosely defined.

---

[1] The principles are not independent: they overlap (e.g., protected variations vs. open-closed principle), and some subsume others (e.g., information hiding vs. encapsulation and abstraction).

In summary, while most researchers agree on what the layered pattern is meant to achieve — reusability, portability, modularity, exchangeability — different authors have proposed different rules to characterize the layered pattern, which may be seen as different ways of achieving the desired qualities of the layered pattern. To tackle the issues outlined above, we carried out a first systematic literature review on existing studies on the layered pattern and software layered architectures. These studies include peer-reviewed papers (found in well-known databases such as Google Scholar, Scopus, IEEE Xplore, and ACM Digital Library), technical reports and software engineering books focusing on the layered pattern. The goal of this first review is to inventory the layering principles reported in the literature and document them in an explicit and uniform manner through rules. In a second systematic literature review, we also analyzed existing software architecture reconstruction approaches targeting the layered pattern in the light of the identified rules.

The audience for both reviews includes researchers, software architects, software developers, as well as technical team leads.

The contributions of this paper are summarized as follows:
- Identification of principles that can help achieve the qualities of the layered pattern
- Translation of the principles into explicit and better-defined rules
- Illustration of these principles using graphical views
- Analysis of reconstruction approaches in relation to these rules
- Identification of some avenues of research resulting from our analysis.

We have structured the remainder of this paper as follows. In Section 2, we present an overview of the layered pattern and we discuss background concepts as well as the motivation behind our review. In Section 3, we present the method we used to carry out both SLRs. In Section 4, we rely on systematic reviews guidelines to perform our first SLR. The latter synthesizes the various rules described in the surveyed studies in terms of more fundamental principles that help us achieve the qualities of the layered architectural pattern. In Section 5, we also rely on these guidelines to present our second SLR. The latter investigates the reliance on these principles to support software architecture reconstruction. In Section 6, we compare our work with related work. In Section 7, we present some threats to the validity of our work. We conclude and outline some perspectives in Section 8.

## 2. BACKGROUND AND MOTIVATION OF THE SYSTEMATIC REVIEW

### 2.1. An overview of the layered pattern

In 1968, Dijkstra [71] laid the foundations for the layered architectural pattern. In 1972, Parnas [72] strengthened the foundations. Numerous books and papers have since presented their own descriptions of this pattern (e.g., [12, 13, 15, 16, 20, 21, 22, 23, 24, 28, 29, 32, 35, 68, 85, 94, 95, 101, 104, 111, 113, 118, 159, 162]). The layered pattern can be defined as: *a technique for structuring a system by decomposing it into sets of tasks [21, 101]*. Each of these sets corresponds to a level of abstraction representing a layer of the system. Each layer uses the services of the lower layer and provides services to its immediate higher layer. The structure of the system is achieved by arranging the layers one above the other, in increasing levels of abstraction. Accordingly, in a well-designed layered architecture, the layers should interact according with a strict ordering relation [22], i.e., a layer may only use the services of the immediate lower layer. This is referred to as *strict layering* in [21, 102, 112, 114, 117, 128], as *strict hierarchy* in [162], and as *closed layering* in [23, 125]. Clements et al. [22] refer to that unilateral restricted ordered relationship between layers as the *allowed-to-use* relationship.

According to Larman [13], the layered pattern consists of organizing "*the large-scale logical structure of a system into discrete layers of distinct, related responsibilities, with a clean, cohesive separation of concerns such that the "lower" layers are low-level and general services, and the higher layers are more application specific.*" In a layered architecture, each layer comprises a set of modules which are therefore cohesive with respect to their responsibilities.

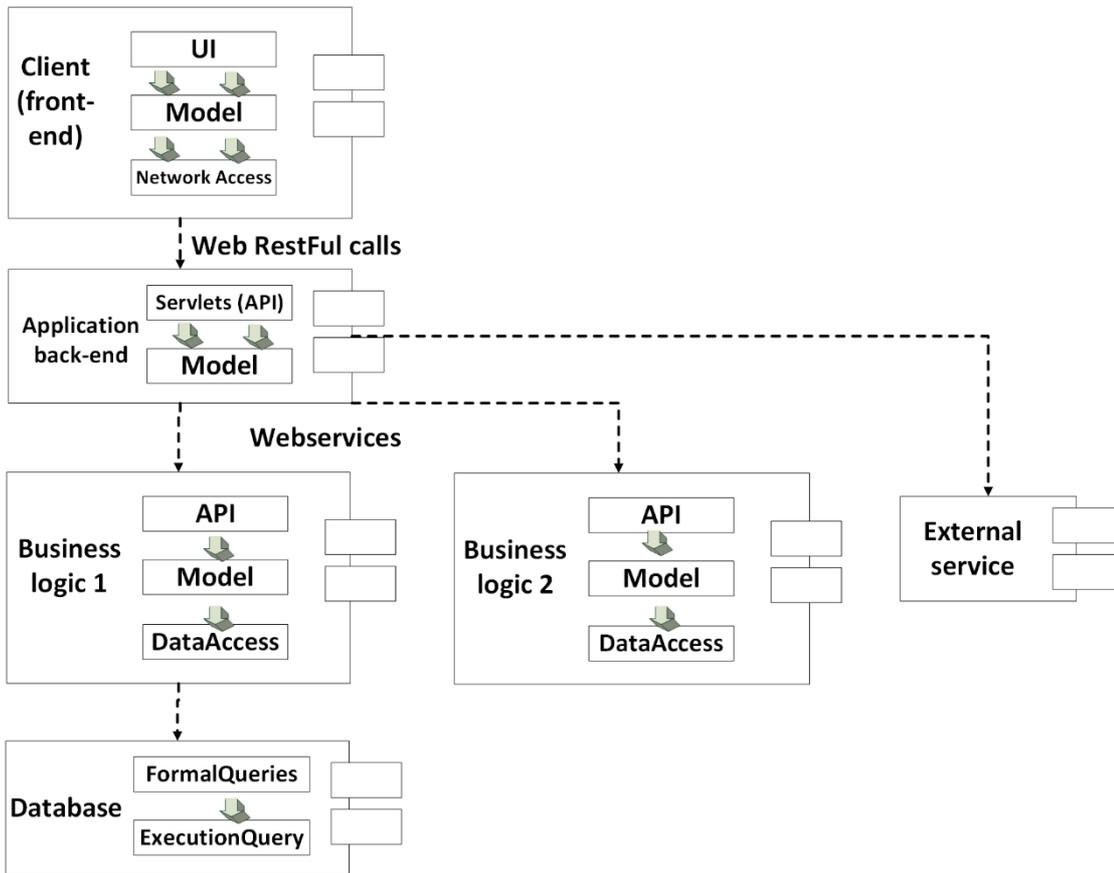

Figure 1. N-Tier components that each are internally designed using layers, but which communicate among themselves using network protocols.

Avgeriou and Zdun [27] explain that in the indirection layer, the logic behind layers is more apparent since a specific layer "hides" the internals of a component or subsystem and offers access to its services. An indirection layer can be integrated to a sub-system as a "virtual machine" or exist as an independent entity. That entity is responsible of redirecting invocations to the sub-system.

Different strategies can be used to partition a software system into layers. The most common strategies are responsibility-based and reuse-based layering [24, 31]. Responsibility-based layering aims at grouping system components according to their responsibilities. Reuse-based layering aims at grouping components according to their level of reuse and assigning the most reusable components to the bottom layers. It is, however, less common than the responsibility-based layering. Examples of these strategies include the OSI (Open Systems Interconnection) model [30] and the e-learning systems [31]. In the OSI model, a layer uses services provided by lower layers and adds value to them to provide services needed by higher layers. Eeles [24] points out that other criteria can be used to define layering strategies: security, ownership, and skill set. Such strategies can be combined to create a multi-dimensional layering. Communication protocols in layers such as OSI and TCP/IP are among the best-known examples of the layered pattern [2]. In particular, the TCP/IP protocol allows transferring data on the internet and comprises four layers, namely: Application (e.g., FTP), transport (e.g., TCP), network (i.e., IP) and data link (e.g., Ethernet [21]). Other examples of layered systems include database systems as well as operating systems [2].

Variants of the layered pattern include *relaxed layering,* also known as *open layering,* i.e., that consists in arranging layers such that each layer is allowed to use services of any lower layer without restriction; and *layering through inheritance* where the layers are arranged through an inheritance relationship [13, 16, 21, 112]. Batory and O'malley [134] discuss the concept of *ad hoc layering*. Systems with such a layering are made of layers resulting from mixtures of functional and object-oriented decompositions. Hence, they do not comply with the definition of a typical layering that is represented as a strict single system-wide stack layering.

As Larman [13] pointed out, originally, the notion of *tiers* and *logical layers* were used interchangeably (e.g., in [24, 29, 31, 88]). Over time, these two notions have evolved and now have very different meanings. A layered architecture resulting from the application of the layered pattern now provides a static view of the system it describes, while an *n-tier* architecture resulting from the application of the *multi-tier* pattern now provides a dynamic view of the system at hand [22] and usually refers to a physical processing node [13]. In more recent research, the layered pattern and the multi-tier pattern should therefore not be confused with each other. For greater clarity: the multi-tier pattern emphasizes run-time distribution of components on different (perhaps virtual) machines, that can often be dynamically reconfigured, such as to adapt to varying loads, and which communicate via networking. The layered pattern more commonly considers software whose layers tend to be brought together at compile time or configuration time, and communicate by procedure calls, although not exclusively. The two can co-exist, with internally layered components becoming tiers in an n-tier system. Figure 1 illustrates that co-existence.

The layered pattern promotes quality attributes such as reuse, portability, maintainability and understandability. But this pattern also comes with liabilities such as potential lack of flexibility and reduced performance [21, 111]. To address these shortcomings (especially in real-time and embedded systems [130]), a current practice is to resort to *relaxed layering* [22, 116], which consists of allowing upper layers to bypass their immediate lower layer to use bottom layers [116, 139]. However, doing this results in what are called *skip-call violations* in [15] and *layer bridging* in [5, 22]. Exceptionally, a layer may need to rely on a service offered by an upper layer. These dependencies are called *back-call violations* in [15] and *upward usage* in [22]. The quality attributes promoted by the layered pattern are no longer supported when layers are allowed to use services of higher layers [22]. This leads to the formation of cyclic dependencies that are likely to make the system monolithic and therefore unbreakable into multiple layers [15]. These situations where there is a cycle between modules assigned to distinct layers are referred to as *cyclic violations* [15, 153]. The structure of a layered architecture must be a directed acyclic graph or at least a directed graph with very few cycles. The empirical study carried out in [42] showed that strict layering is not often enforced, since skip-calls are made extensively, whereas there are rarely any back-calls.

As Kazman and Carrière [5] point out, there is no agreed way to enforce a "layer" when implementing a system since there is no explicit "layer" construct in any modern programming language. Layering may indirectly be achieved by relying on programming language constructs (e.g., modules or inheritance) with accessibility constraints. Such programming constructs help convey the layering, but do not allow enforcing it. Still, means such as naming conventions, code ownership, and design conventions can be considered as attempts to enforce layering.

## 2.2. Motivation for carrying out both SLRs

The surveyed theme is topical for the targeted audience for several reasons:
- *Software architecture is important*: It is the most important abstraction generated during the software lifecycle. Thus, being aware of architectural patterns is essential to properly structure software systems. The layered pattern is one of the most common architectural "tools" used to structure software systems [70, 18].
- *Architecture reconstruction is important to reengineer legacy systems without losing the domain knowledge embodied in them*: Legacy software systems embed important knowledge acquired over the years, are of a proven technology, which makes them critical assets for enterprises [34, 79, 165]. Several billion lines of legacy code exist, yield high maintenance cost, are prone to failures due to lack of experts and suppliers/vendors [79], and an important portion of them comply with the layered pattern. That code needs to be evolved toward new technologies to enable progress in business practices [165]. That evolution requires understanding the code, and the architectural principles used to lay it down.
- *The layered pattern is a prevalent architectural pattern*: A recent study carried out by Muccini and Moghaddam [155] showed that the layered pattern is the most used architectural pattern to structure IoT (Internet of Things) systems. Pahl et al. [44] explain how layering can be used to achieve service-orientation. As such, layering is therefore very useful for the realization of services. There is thus a need to extensively discuss the usage of layering principles. In this paper, we discuss such usage in the context of the reconstruction of software layered architectures.
- *The layered pattern has not been defined* consistently: There is a lack of uniformity in the literature discussing the layered pattern [70, 105]. The resulting architecture descriptions may often clash. In student projects for instance, several designs of layered systems usually focus on the description or illustration of the layers and of their names, without specifying the contents of such layers, the communication rules they use, or even their rationale [105]. As such, the resulting architectural product provides little guidance to the developers, while the terminology used to designate the corresponding layered pattern may be extremely confusing [105]. Furthermore, papers are usually vague about the essence of layering principles [70]. There is therefore a need to revisit the layering principles that the layered pattern conveys to make their essence more precise and uniform

- *The layered pattern architecture reconstruction serves several purposes*: The layered pattern features (e.g., architectural principles and violations) can have several uses [4, 8, 143, 163], ranging from architecture enforcement [158], the design of software architectures, their analysis, their reconstruction, their migration, their co-evolution, their restructuring, to the detection of architecture erosion (e.g., using tools such as SonarGraph [120, 161, 163, 166, 168]). In this paper, we focus on the reconstruction of software architectures. In particular, the reconstructed architecture serves as a medium to conduct diverse analyses such as pattern conformance, dependency analysis or quality attribute analysis [4]. These analyses can help identify technical debt –that is extensively discussed in [147, 148, 149, 154, 157] – and more particularly architectural technical debt (ATD). ATD is very hard to identify [90]. It is the most toxic dimension of technical debt, and the most challenging to uncover due to the lack of tools to identify and manage it [99, 120].

## 3. METHOD USED TO REVIEW THE LITERATURE

### 3.1. Research questions

To analyze the literature on the layered pattern, we performed two systematic literature review (SLRs), by following the guidelines on systematic literature reviewed proposed by Kitchenham [169]. Our two SLRs respectively aimed at answering the following research questions:

- **(RQ1)**: What are the fundamental layering principles and the rules they encompass? This question aims at: 1) Identifying the principles that can help achieve the qualities of the layered pattern; 2) Translating these principles into explicit and better-defined rules; and 3) Illustrating these principles using graphical views. We investigate this question in Section 4.
- **(RQ2)**: Which of these principles and rules have been used to reconstruct software architectures compliant with the layered pattern, and how? This question aims at analyzing reconstruction approaches in relation to these rules. We investigate this question in Section 5.

The investigation of both research questions helped us synthesize the layered architectural pattern principles discussed in the literature and explore the use of their rules in the software architecture reconstruction. That investigation also helped us outline some avenues for future work (see Section 8).

### 3.2. Identification of the research

For each of our SLR, we started with some known and respected literature and comprehensively searched for other literature by looking for papers, while also using some keyword searches based on numerous important terms in the field. To identify the search, we therefore relied on two search strategies: the database-driven strategy as well as the manual search strategy. In particular, the database-driven strategy is based on searching specific databases using concise, well-defined, and repeatable queries. Noteworthy, the manual search and database-driven strategies are the most used search strategies to identify primary studies [170]. To ensure the completeness of the search, we could have used additional search strategies (e.g., snowballing), but due to time-constraints purposes, we chose to focus on these two strategies.

To carry out the database-driven search, we started by searching six well-known databases, namely: Google Scholar[2], ScienceDirect[3], ISI Web of Knowledge[4], ACM digital Library[5], IEEE Xplore Digital Library[6], and Scopus[7]. Our search was based on sentences made of keywords belonging to the following groups:

**Group 1**: Reconstruction, recovery, discovery, analysis, study, design
**Group 2**: software architecture, software system
**Group 3**: layers, layered, layering, hierarchy
**Group 4**: rules, principles.

---

[2] https://scholar.google.com/
[3] https://www.sciencedirect.com/
[4] https://www.webofknowledge.com/
[5] https://dl.acm.org/
[6] https://ieeexplore.ieee.org/Xplore/home.jsp
[7] https://www.scopus.com/

We searched the primary studies in databases by composing keywords of the four groups with the AND coordinating conjunction and keywords inside each group with the OR coordinating conjunction. We have obtained these keywords by iteratively refining a list of keywords relevant for the two SLRs so as to create a query that included papers that seemed relevant (based on their titles). The initial query resulting from this process is the following:

*Query = (Reconstruction OR Recovery OR Discovery OR Analysis OR Study OR Design) AND ("Software Architecture" OR "Software System") AND (Layers OR Layered OR Layering OR Hierarchy) AND (Rules OR Principles)*

This allows querying references from the selected databases to download studies/references. Obviously, we sometimes needed to adapt the initial query specified above from a database to another since some of these databases have their own algorithms to search the queries that they structure accordingly. Note that we performed a basic default search in most databases by just typing the query(ies) as-is in the search box of the database.

However, in a few databases, we can only use our query in a customized search. Hence, in IEEE for instance, we typed the adapted queries for this database in the command search box which is an advanced search option offered by this database.

Google Scholar allows exporting references one by one, which can be a very tedious task, especially when running a query on that database returns hundreds or thousands of potential primary studies. Hence, to directly export all the resulting references in an EndNote-like format, EndNote being our reference manager, we used the Publish or Perish[8] tool where we ran the query adapted to Google Scholar. Hence, we typed the query in the "All of the words"[9] pane of the Google Scholar Query panel available on the Publish or Perish tool.

To select papers to include in our first SLR, we relied on a set of inclusion and exclusion criteria. As such, we included a study in the first review if it matched all the following inclusion criteria:

- It is a conference paper, a magazine, a workshop paper or a journal paper;
- It has been peer-reviewed;
- It targets the computer science field;
- It proposes techniques to reconstruct, analyze or design software architectures compliant with the layered pattern;
- It conveys architectural rules/principles that can be used to perform the reconstruction, analysis or design of software architectures.

We excluded from the review studies that matched any of the following exclusion criteria:

- It is not published in English;
- It does not describe the layered pattern's architectural rules or specificities;
- It is a survey paper only;
- It considers the layered pattern as the multi-tier pattern whereas these are now considered as two distinct concepts (see difference in Section 2.1);
- Its citation (reference) or abstract is available but not its full content;
- The full paper is not available due to a retraction notice (e.g., because it is violating IEEE's publication principles) or because of copyrights restrictions preventing a database (e.g., ACM) from providing the full content of the document.

We performed the following rounds to select the primary studies that will be included in our review:

1. Construction of a reference list in the reference manager EndNote by searching the queries in the chosen databases and exporting them in an Endnote-like format (e.g., .ris).
2. Removal of references with no title (i.e., empty Title field) or no author (i.e., empty Author field).
3. Elimination of duplicates in the reference list by using EndNote.
4. Reading the titles and then the abstracts of the papers so as to remove papers that do not match the inclusion and exclusion criteria.
5. Reading of the introductions and conclusions to eliminate irrelevant papers according to the inclusion and exclusion criteria.

---

[8] https://harzing.com/resources/publish-or-perish.
[9] https://harzing.com/resources/publish-or-perish/manual/using/data-sources/google-scholar?source=pop_6.33.6259.6749

6. Brief reading of the paper's text to eliminate additional papers based on the inclusion and exclusion criteria.
7. Manual adding of documents that we deem relevant for the survey and that where possibly missed when carrying out the previous selection rounds. These were papers that we are aware of as experts in the field, but for which we could not adapt keyword searches such that they would be found. In particular, we manually added additional primary studies that we deemed relevant for the first part of the SLR. Note that at this step, we also considered one additional category of documents i.e., technical reports that we deemed relevant for the SLR.

The manual search and the database-driven search resulted in 83 primary studies that we included in our first SLR. Note that we discarded potential primary studies treating tiered architectures as layered architectures (e.g., Scanniello et al.'s studies i.e. [29] and [88]). Still, we have not discarded the paper of Paris [31] since, even though she is assimilating the classical responsibility-based layering strategy to the 3-tier pattern, she is also discussing a different layering strategy (i.e., reuse-base layering) that is orthogonal to the classical one. And since our SLR is not a tertiary study, we also discarded potential primary studies that were surveying a related theme (e.g., Ducasse and Pollet [4], Savolainen and Myllarniemi [70], and [105]).

To select studies for the second systematic literature review, we reused the same list of potential primary studies obtained in the first round of the first the SLR through the database-driven search. But we only focused on studies using the layered pattern principles to reconstruct software architectures. Hence, we had to slightly modify the initial inclusion and exclusion criteria specified above. We therefore discarded primary studies considering tiered architectures as layered architectures (e.g., [29], [65] and [88]). Hence, the aforementioned selection process resulted in 37 primary studies that we included in our second SLR. Since we are not performing a tertiary study, we have not considered certain studies that also present surveys, but for a related theme (e.g., [4]). Wong et al. [94]'s paper was a potential paper that was obtained through the database-driven search. But, even though it discusses the layered pattern specificities as well as introduces a reverse engineering approach, we have not included that work in the second SLR. The rationale is that the DRH (Design Rule Hierarchy) layers it attempts to reconstruct are conceptually different from the typical layered pattern's layers.

Note that, to make the search of primary studies more manageable, we have excluded books from the database-driven search process. Still, we have manually included a couple of software engineering that are well-known and that describe fundamental aspects of the layered architectural pattern.

Table 2 in the Appendix lists the primary studies respectively included in the first SLR, in the second SLR, or in both SLRs.

### 3.3. Strategy to extract and synthesize data

An iterative analysis of the primary studies included in the first SLR led us to the identification of four principles to be followed when applying the layered pattern:
- *Abstraction*,
- *Responsibility*,
- *Transversality*, and
- *Protection against variations*.

From these principles, we derive and detail a catalogue of six layering rules, which is more exhaustive than what has been published by previews secondary studies on the layered pattern (e.g., [70]). These rules are formulated based on the information we retrieved from the primary studies, and they embody the fundamental design principles conveyed by the layered pattern. These rules are:
- The *Layer Abstraction Uniformity* rule,
- The *Incremental Layer Dependency* rule,
- The *Responsibility rule*,
- The *Transversality* rule,
- The *Interfacing* rule, and
- The *Stability* rule.

Figure 2 depicts the layering principles as well as the six related rules, i.e., design rules they encompass. It also depicts the relationships between them. We further discuss these principles and rules in the next sections.

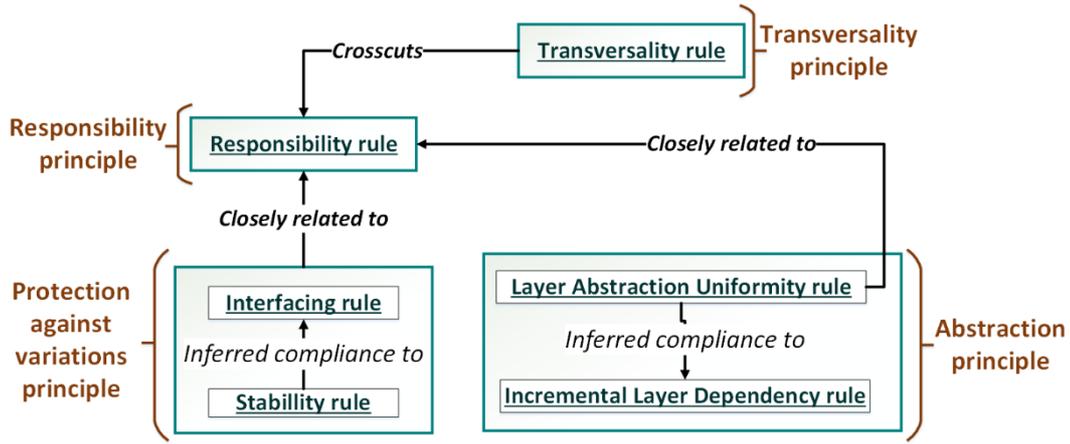

Figure 2. Catalogued layering rules and relationships between them.

We relied on these principles and rules to extract data from the primary studies included in our first SLR. We then extracted data from the primary studies included in the second SLR in the light of these principles and rules. In the following, we report the synthesis of the extracted data. Thus, based on these data, we present in section 4, a synthesis of the fundamental principles and rules characterizing layers within legacy applications, and we group them under four principles that embody the fundamental design principles underlying the pattern: this constitutes our first SLR's results. In section 5, we rely on these rules to synthesize the literature on software architecture reconstruction: this constitutes our second SLR. Section 4 therefore provides the foundations we used to carry out the second SLR reported in Section 5. In this regard, Section 4 therefore establishes the main criteria characterizing the layered pattern based on the current literature on the layered pattern fundamentals, while Section 5 uses these criteria to analyze the current literature focusing on the software architecture reconstruction.

## 4. (RQ1) Results of the first SLR: on the Architectural layering principles

In this section, we present approaches that focus on architectural principles used to design, analyze or reconstruct software architectures compliant with the layered pattern. Violating these principles notably leads to a non-uniform layered design since it encompasses structural violations. Such a design does not comply with the design rules conveyed by the layered pattern and, therefore, results in ATD accumulation [146].

To summarize our findings in terms of principles, we have adopted the guidelines of Systematic Literature Reviews [169] for finding and aggregating the data found in the documents included in the review (i.e., primary studies). The presented principles come from the following 83 primary studies: [2, 8, 13, 15, 16, 18, 19, 20, 21,22, 24, 25, 26, 27, 28, 30, 31, 33, 37, 39, 40, 41, 42, 43, 47, 57, 60, 63, 64, 71, 72, 74, 86, 89, 94, 95, 96, 98, 100, 101, 102, 103, 104, 105, 106, 107, 109, 110, 111, 113, 114, 115, 116, 117, 118, 121, 123, 124, 125, 126, 127, 128, 129, 130, 131, 133, 134, 135, 136, 137, 138, 139, 140, 141, 142, 144, 145, 151, 152, 153, 156, 162, 171]. These studies include: 1) peer-reviewed papers that we retrieved either from well-known databases or from the manual search, and that we retained because they explicitly discuss some layering principles; 2) technical reports; 3) and a couple of software engineering books focusing on the layered pattern.

In the following, we synthesize the data extracted from the primary studies (included in the first SLR) to describe each layering principle and related rules embodying the essence of the layering pattern.

### 4.1. The abstraction principle and related rules

Applying the layered pattern consists of partitioning the system into a set of layers ordered according with the abstraction criterion that governs the flow of communication between components. Hence, the layered pattern allows structuring a system by decomposing it into sets of tasks (e.g., [101]). Each set of tasks corresponds to a given level of abstraction and represents a layer of the system. Each layer uses the services of the lower layer and provides services to its immediate higher layer (e.g., [13, 64, 116, 140]). The structure of the system is then achieved by arranging the layers one above the other, in an increasing level of abstraction. This is consistent with:

- The "**skip-call principle**" (e.g., [15, 152, 153])—that is similar to the "*skip-call ban*" rule (e.g., [43] and [113]), the "*Forward (Skip) Level Invocation*" rule [118], and the "*Is not allowed to skip-call*" rule (e.g., [ 96]). The "*skip-call principle*" states that "*Each layer should aim to maximize the dependency with the layer immediately below and aim to minimize the dependencies with the layers further down.*" [15].
- The "**back call principle**" [15, 152, 153]—that is similar to the "*back call ban*" rule (e.g., [43] and [113]), the "*Backward (skip) Level Invocation*" rule [118], and the "*Is not allowed to back call*" rule (e.g., [96]). The "*back call principle*" states that "*a given layer should not depend on any of its upper layers*" [15].
- The "**cyclic dependency principle**" [15]. The Cyclic dependency principle states that: "*if there exists a set of modules that are cyclically dependent, such modules must belong to a single layer*" [15]. This principle is related to the "*Principle of Minimization of Cyclic Dependencies Amongst Modules*" and to the "*Principle of Maximization of Unidirectionality of Control Flow in Layered Architectures*" [145].

In accordance with the increasing abstraction criterion, layers are usually represented as stacked rectangles, thus yielding a stack diagram [25]. The number of layers to represent usually varies in the literature (see Section 2.1). Figure 3 illustrates the architecture of a layered system as conveyed by the abstraction principle. This figure shows the directions of the dependencies between layers, and therefore enables distinguishing potential violations (e.g., back-calls and skip-calls) from allowed dependencies and potential intra-calls. The literature often depicts such violations as part of layered architectures (e.g., [15, 16, 26, 39]). In Figure 3, the thick arrows represent allowed dependencies. The dashed arrows represent violations: the skip-calls and back-calls are respectively depicted in blue and red. The long-dotted line arrows represent intra-calls. The strength of the connections between two modules may be indicated on the arrow using a numerical value, i.e., a weight.

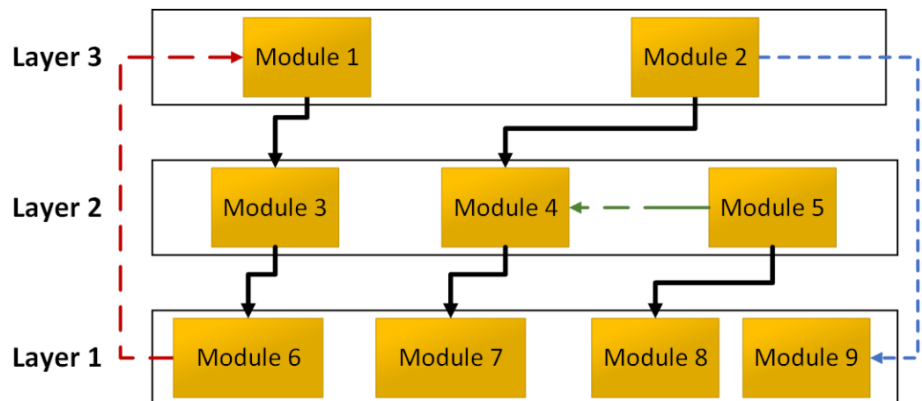

*Figure 3. Illustration of the abstraction principle with violations and intra-calls (adapted from [15, 16, 26, 39]).*

The abstraction principle encompasses two fundamental architectural rules that should guide the design as well as the reconstruction of layered architectures: the *Layer Abstraction Uniformity* rule and the *Incremental Layer Dependency* rule.

*4.1.1. The Layer Abstraction Uniformity (LAU) rule*

This rule states the following (e.g., [33, 37]):

> *Components of the same layer must be at the same abstraction level so that the layer has a precise meaning*.

The level of abstraction of a component often refers to its conceptual distance from the "physical" components of the system [21], i.e., hardware, database, files and network. Components at the highest levels are application-specific; they generally contain the visible functionality provided by the system.

An important concern when designing a layered architecture is to find the right number of layers, i.e., abstractions. This requires making compromises: a high number of layers introduces an unnecessary overhead, whereas a low number may lead to a monolithic system [21]. Most layered systems have three layers (usually the presentation, business and data layers); a few exceptions include the 4-layered architecture discussed by Bouihi et al [100], which reengineers the traditional layered architecture of current e-learning and m-learning platforms by adding a semantic layer. When discussing the *development view* that is concerned with the organization of modules in a system, Kruchten [63] recommends defining four to six layers when designing a system in a layered fashion. This is consistent with Kouamou and Kungne [156] who state that enterprise information systems should consist of four layers. This is also consistent with Mohammed and Fawcett [162] who describe a four-layer general layering.

*4.1.2. The Incremental Layer Dependency (ILD) rule*

This is related to the "ideal layering" property that states (e.g., [8, 26, 33, 64, 107, 110, 117, 121, 123, 124, 127, 135, 141, 144]):

> *A component in a layer (j) must only rely on services of the layer below (j-1).*

Even though enforcing The Incremental Layer Dependency rule is a basic assumption of any layer-based architecture, that rule is the most violated to foster system performance and flexibility either through back-calls (to higher layers), or skip-calls. Zdun and Avgeriou diverge from that rule by not forbidding back-calls, arguing that it is a specific issue to the layered pattern realization [127].

It is worth pointing out that there is no clear consensus among researchers on the acceptability of intra-dependencies i.e., dependencies between components in the same layer. Such dependencies are accepted by some (e.g., [25, 27, 28, 63, 64, 111, 127, 142, 162]) and not recommended by others (e.g., [20, 41, 113]). Our analysis of the various descriptions of the layered pattern and several open-source systems led us to conclude that acceptance of intra-dependencies depends on the granularity of the components (e.g., packages) of the layer: the higher the granularity, the lower the number of intra-dependencies there should be [18]. The Incremental Layer Dependency rule should thus be stated in a way that allows intra-dependencies and the skip-calls and – if absolutely necessary to reflect existing systems – back-call violations. We therefore rephrase this rule as follows:

> *Components of layer j-1 should be designed to primarily offer services to components of layer j.*

This means that, for a given layered system: (1) the most desirable dependencies are downward dependencies between adjacent layers, (2) skip-calls are generally accepted, (3) intra-dependencies are not recommended but may be tolerated, and (4) back call dependencies are to be avoided. We derived this rule not only from our analysis of various descriptions of the layered pattern (e.g., [25, 64, 85, 95, 110, 128, 159]) as done for other rules, but also from our analysis of several open-source systems (e.g., [18, 33, 37]). Note that this rule is also consistent with the empirical study described in [42], which has shown that strict layering is generally not enforced in layered systems.

Compliance with the first abstraction rule (Layer Abstraction Uniformity) implies that the packages of the same layer should be at the same distance from the "physical" components of the system. However, the presence of back-call and skip-call dependencies causes a discrepancy between the packages' distances, even when they belong to the same layer. Thus, compliance with the first rule derives largely from compliance with the second rule (i.e., Incremental Layer Dependency).

**4.2. The responsibility principle and rule**

Clements et al. [74] define the responsibility as: "*a general statement about an architecture element and what it is expected to contribute to the architecture. This includes the actions that it performs, the knowledge it maintains, the decisions it makes or the role it plays in achieving the system's overall quality attributes or functionality.*"

Applying the layered pattern implies decomposing the system into a set of cohesive components (i.e., responsibilities) and properly assigning these components to a set of abstraction levels. We refer to this as the responsibility principle, which is related, in this context, to the abstraction principle. The responsibility principle encompasses a single rule, namely: the responsibility rule. This rule states (e.g., [21, 33, 37]) the following:

> *Each layer of the system must be assigned a given responsibility so that the topmost layer corresponds to the overall function of the system as perceived by the final user, and the responsibilities of the lower layers contribute to those of the higher layers.*

The responsibility rule is consistent, to some extent, with the responsibility convention rule [43]. The latter states that: "*All elements of the module must adhere to the specified responsibility*". A module can be, for instance, a subsystem or a layer [43].

In a given layer, each responsibility is implemented by a set of interacting components that need to be cohesive and specific to a given domain (e.g., [8, 24, 25, 26, 27, 128]). Therefore, each component of the system should be designed to implement a specific service and must belong to a single layer. Each layer therefore comprises related responsibilities (e.g., [43, 113, 116, 126]). By promoting the interaction within components of the same layer, this rule implicitly favors intra-calls especially when the granularity of the layers' component is quite fine-grained. Note that different books further describe the essence of the Responsibility rule (e.g., [1], and [13]). In particular, the concept of responsibility is defined by Bass et al. [1] as "*the functionality, data, or information that a software element provides*". Thus, the logic of a software system is divided into several responsibilities, each layer grouping similar responsibilities [26]. A given responsibility can be singular or constituted of several finer-grained responsibilities [39]. To reduce the complexity and increase the comprehensibility of a given layer, that responsibility can be recursively refined until basic responsibilities are obtained [39].

Figure 4, adapted from [24], shows an example of a processing system organized around three concepts: *Customer*, *Order*, and *Product*. That system is layered according with three main responsibilities isolated from each other and respectively implemented by a layer: the *presentation logic*, the *business logic* and *the data access logic*. Hence, the presentation logic layer has three components: *CustomerView*, *OrderView* and *ProductView*. These components are respectively responsible for the *presentation logic* associated with the *Customer*, the *Order* and the *Product* (e.g., rendering of a customer in the user interface). Likewise, the *business logic* layer has three components: *Customer*, *Order* and *Product*. These components are respectively responsible for the business logic associated with the *Customer*, the *Order* and the *Product* (e.g., validation of *Customer* details). The data logic layer has three components: *CustomerData*, *OrderData* and *ProductData*. These components are respectively responsible for the *data logic* associated with the *Customer*, the *Order* and the *Product* (e.g., persistence of the *Customer* information). Note that the number reported on a given arrow going from a component *A* to a component *B* indicates the number of times the entities (e.g., classes or packages) of component *A* use the entities of component *B* (e.g., through method calls, data accesses).

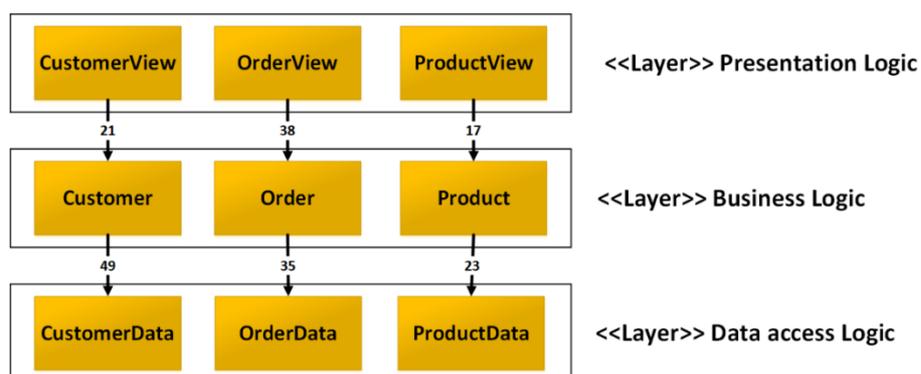

*Figure 4. Illustration of the responsibility principle (adapted from Eeles [24]).*

### 4.3. The transversality principle and rule

By convention, software layers are generally drawn horizontally as thin boxes that are stacked on top of each other with the physical medium at the bottom. With such an alignment, components that are used by many other components, such as libraries, are usually assigned to lower layers. In practice, such an assignment goes against the abstraction principle and particularly against the Incremental Layer Dependency rule since it can induce many skip-calls. Therefore, the transversality principle recommends assigning the most-used components to transverse components (i.e., vertical boxes); this resolves that issue by turning the skip-calls generated by these components into adjacent dependencies directed from the horizontal layers' components to the vertical (transverse) ones. These are called omnipresent components in [8, 18, 47]. They are also called library components (e.g., [18]). The transversality principle encompasses the transversality rule which stipulates the following:

> *Components that are intensively used by components in different horizontal layers should be placed in a transverse component instead of a horizontally oriented one.*

The component resulting from the enforcement of this rule should therefore be aligned perpendicularly to the horizontal layers. Such an arrangement of layers is notably advocated by Bachmann et al. [25] through the concept of "layers with a sidecar" that is common in many layered architectures (e.g., the reference layered architectures proposed in [101, 136]). Bourquin and Keller [20] also advocate in favor of a lateral layer. That layer is respectively called vertical layer in [18] and transversal layer in [2].

Figure 5 illustrates the transversality principle in case of a four-layer architecture. The figure has either one of the two following meanings (or both in case of a poor layered architecture) [25]: (1) modules in layer 1 are allowed to use modules in layers 2, 3 and 4, or (2) modules in layers 2, 3 and 4 are allowed to use modules in layer 1. It is up to the designers of the depicted architecture to specify the intended meaning.

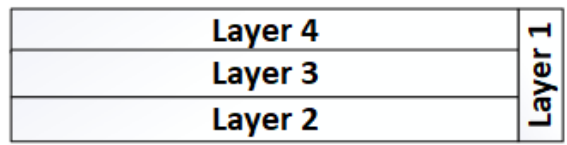

*Figure 5. Illustration of the transversality principle (adapted from Clements et al. [22] and Bachmann et al. [25]).*

### 4.4. The protection against variations principle

The protection against variations principle states*: "each layer must be designed so that the variations affecting its inner entities have no adverse impact on other layers"*. This principle is strongly inspired by the notion of *protected variations*, which is further discussed in some books (e.g., [13]). It also relates to some extent to *information hiding* and to the *separation of interface and implementation* design principles (e.g., [95]), and to the visibility convention (e.g., [43, 113]). In particular, information hiding and encapsulation simplify the analysis of a system [58] because they support the description of modules without the need to explain their implementation details [58]. The protection against variations principle encompasses the two following rules: the interfacing rule and the stability rule.

The protection against variations rules are closely related to the responsibility rule and to some extent to the abstraction principle. In particular, when reconstructing software layers, the responsibility rule may help identify cohesive components/modules while the interfacing rule may help identify which of these components share the same interfaces.

#### 4.4.1. The interfacing rule

This rule states that:

> *A layer should provide an input interface that can be used by the layer immediately above it. Here, we consider that the term interface does not necessarily refer to a given API but rather to the medium that allows accessing a given layer.*

Thus, the input interface of a given layer is made of the entities – classes or packages depending on the granularity on which we reason – through which its services are called by the upper layer. An interface could therefore be seen as a fence whose gate allows centralizing the accesses to the inner implementation of a layer (this may favor some intra-calls in layers). According with the interfacing rule, a layer should hide, as much as possible, its internals using an interface. The latter will allow the limitation of the access points to a given layer's services [26]. Therefore, the more this rule is enforced, the less coupling there is between the layers. Thus, the impact of modifying a given layer is reduced in other layers and as a result, the maintainability and the portability of the system are increased. The use of interfaces to hide the layers' implementation details is a practice advocated by several authors (e.g., [21, 26, 30, 106, 110, 116, 117, 133, 151]) who usually recommend standardizing them (e.g. [2, 26, 30]) and making them public (e.g., [60, 89, 127]). The Facade pattern [59] and the Delegation Adaptor primitive [139], can be used to implement the interfaces of layers [21, 60]. By being a means to get rid of complex or circular dependencies, the Façade pattern help layer a system [59]. Besides, the Delegation Adaptor primitive is a means to separate the explicit interface of a layer from its implementation [139].

Buschmann et al. [21] proposed different approaches to specify the interface of each layer based on the extent to which a layer and its internals are allowed to interact with other layers. These include [21]:

- **Black-box approach**: each layer $j$ is a black box for layer $j+1$. Layer $j$ therefore offers its services through a flat interface that may be encapsulated in a Façade object. This approach supports a strict enforcement of the Interfacing rule.
- **Gray-box approach**: a layer $j+1$ is not able to see the internals of layer $j$'s components, but layer $j+1$ knows the number of components that layer $j$ comprises and can therefore interact with each of them. This approach supports a relaxed enforcement of the Interfacing rule.
- **White-box approach**: a layer $j+1$ is aware of the internals of layer $j$. This approach therefore means not enforcing the Interfacing rule at all.

The gray-box approach is a compromise between the black-box and the white-box approach [21]. Of particular note, the *Interface violations* proposed by Bischofberger et al. [41] can be considered to some extent as violations of the *Interfacing rule*. An interface violation occurs when some subsystems use the non-interface artifacts of some subsystems whereas it is not allowed in [41].

Some of these three approaches can be used to enforce the Protection Against Variations principle, and particularly the Interfacing rule. As such, they lead to an x-box diagram that depicts a layered software architecture. In this context, x= black or gray, depending on the extent to which the Interfacing rule is enforced. Figure 6 adapted from Buschmann et al. [21] shows an illustration of the gray-box approach (in this case, x=gray) to the Interfacing rule. In this figure, each layer comprises three components. Components belonging to the same layer or to distinct layers call each other directly. For instance, in the middle layer, two components interact directly, i.e., *Component_2.2* makes a call to *Component_2.3*. In addition, *Component_2.1* from Layer 2 calls *Component_1.1* and *Component_1.2* of Layer 1 directly.

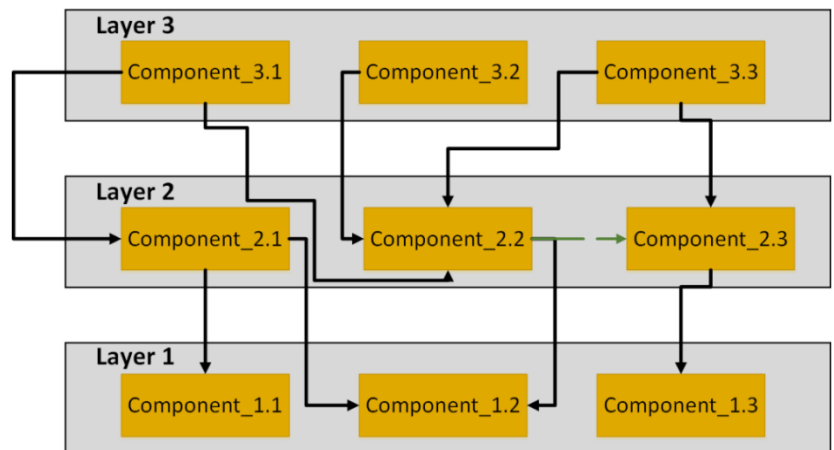

*Figure 6. Gray-box approach to the Interfacing rule illustration (adapted from Buschmann et al. [21]).*

Figure 7 illustrates the black-box approach: each layer offers its services through a unified interface (i.e., represented by a green rectangle); components belonging to the same layer can still call each other directly whereas calls between components belonging to distinct layers goes through the unified layer interfaces. For instance, *Component_2.1* calls a Layer 1 interface object (i.e., *Int1*) that forwards the request to *Component_1.1* (or *Component_1.2*).

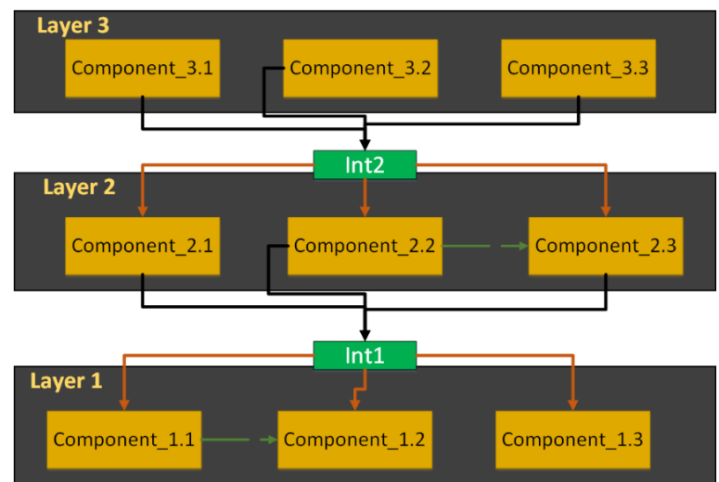

*Figure 7 Black-box approach to the Interfacing rule illustration (adapted from Buschmann et al. [21]).*

**4.4.2.** *The stability rule*

This rule stipulates the following:

> *The interfaces of each layer should not change if the functionality accessed through the layer interface has not changed (even if the implementation behind it has changed).*

This means that if the implementation of a layer were to change, that layer should continue to provide the same services to its clients [60, 26, 27] and a client of this layer should be able to keep using it [2] without having to adjust its own interfaces to cope with the new layer's implementation. The more this rule is enforced, the less the modifications of a given layer affect its users. Thus, this rule not only allows increasing the reuse of a system [2] but also reduces the maintenance cost and effort [61]. The literature (e.g., [26]) hints that compliance to the Stability rule derives from compliance with the Interfacing rule since the stability of a layer can notably be supported using input and standardized interfaces. This stability may be further enhanced using an output interface that would prevent the changes in this layer to ripple down to its lower layers.

## 5. (RQ2) RESULTS OF THE SECOND SLR: ON THE USE THE LAYERED PATTERN PRINCIPLES TO RECONSTRUCT SOFTWARE ARCHITECTURES

The analysis of the primary studies included in the second SLR shows that the reconstruction process often focuses on the generation of an architecture that complies with a specific pattern to reconstruct the architecture of a legacy software system. Figure 8 adapted from [8] shows the architecture reconstruction process driven by the applied architectural pattern. Accordingly, a pattern-based architectural reconstruction is usually performed in four steps:

1. **Step 1**: The analysis of the architectural pattern to extract its fundamental principles. As done in Section 4, a synthesis of the literature describing the architectural pattern can help extract such principles
2. **Step 2**: The use of a selected set of extracted principles and of the set of the constraints they convey to define a reconstruction approach(es). The latter aims at abstracting the system's information (facts) using these principles.
3. **Step 3**: The extraction of the information from a legacy system. Such facts can for instance be structural or lexical facts retrieved from the analysis of the system's source code.
4. **Step 4**: The reconstruction of higher-level models using the right abstraction technique compliant with the architectural pattern's principles and constraints. That technique abstracts the information facts to generate a reconstructed architecture.

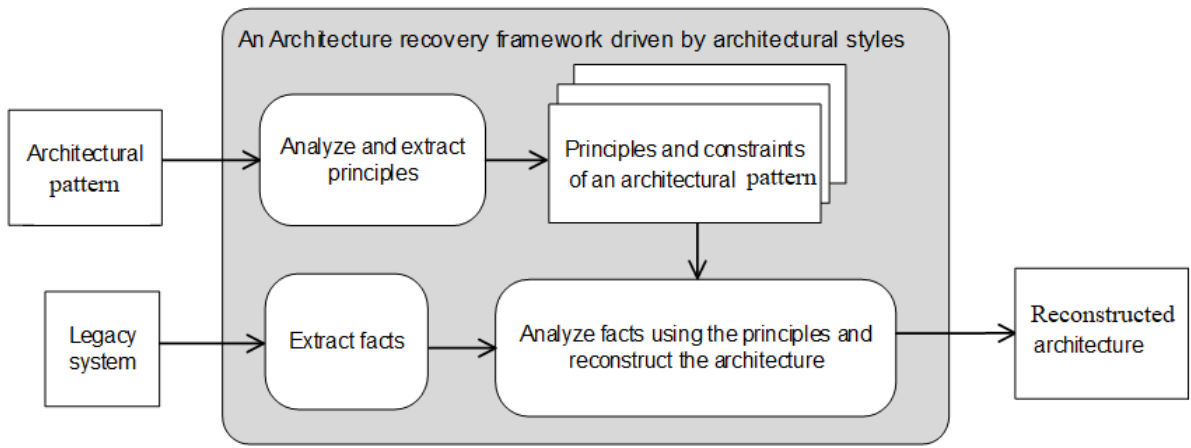

*Figure 8 Reconstruction process driven by the applied architectural pattern (adapted from [8]*

Existing approaches to software architecture reconstruction exploit the principles of the applied pattern in different ways and to various degrees. In the following, we discuss the use of the layering rules in software architecture reconstruction approaches. Our focus is on their use of the rules and their degree of automating the use of these rules in reconstruction. Table 1 reports the 37 primary studies on which our analysis focuses. These studies consist in 37 peer-reviewed papers that we retrieved from well-known databases and from the manual search, and that we retained because they rely on layering principles to reconstruct software architectures.

### 5.1. On the enforcement of layering rules in the reconstruction process

As Table 1 illustrates, different approaches used different layering rules as input to reconstruct software layers. In the following, we present the variety of rules they used, the combination of such rules during the reconstruction process, and the tolerance of some deviations from these rules during the reconstruction.

*5.1.1. On the variety of rules used in the reconstruction process*

Based on the rule usage that Table 1 reports, we can classify layering rules used to reconstruct software architectures according to three categories of rules:

- **Core rules**: Approaches supporting layered architecture reconstruction systematically use abstraction rules (i.e., LAU and ILD rules) when reconstructing layered systems. More than sixty percent of these approaches (e.g., [15, 20, 21, 28, 102, 40, 21]) rely on the responsibility rule to recover responsibilities before assigning them to abstraction levels.
- **Optional rules**: Only a few of these approaches (e.g., [8, 20, 64, 129]) rely on the transversality rule. This might be because some may consider the transversality rule as being a design decision, and not a clear rule to follow when designing or reconstructing layered architectures. Furthermore, the analysis of these approaches shows that the number of reconstruction approaches relying on the Interfacing rule is quite limited.
- **Never used rules**: None of the surveyed approaches relied on the Stability rule when reconstructing software layers. The inclusion of this rule therefore distinguishes our catalogue of rules from existing work. Using this rule for reconstruction is challenging as it requires reconstructing architectures of different versions of the system and using appropriate metrics and methods to compute interface similarities.

*5.1.2. On combining multiple rules for reconstruction*

Very few approaches support more than three out of the six catalogued layering rules. The few that diverge from that trend include [8, 47, 57, 64, 121, 129, 137]. Surprisingly, most of the approaches (e.g., [47, 57, 129]) that cover most of the catalogued rules, are the ones supported by the Rigi tool that was developed more than two decades ago by Müller et al. [47]. This suggests that the layering reconstruction field has somewhat relaxed over the past years, preferring to focus only on the core layering rules, i.e., the abstraction rules and to some extent the responsibility rule.

However, not enforcing one or many layering rules when reconstructing software layers may lead to some issues. For instance, not enforcing the responsibility rule and relying instead on the existing decomposition of systems into components (e.g., packages) while reconstructing software layers may lead to the dispersion of cohesive responsibilities throughout the system's layers as pointed out in [39]. For example, when we flatten the component hierarchy, two child components that are nested within the same parent component might be, during the reconstruction process, assigned to distinct layers to enforce abstraction rules. These components may therefore end up, wrongly, in two distinct adjacent layers. Many layering reconstruction approaches are exposed to this issue (e.g., [16, 19]).

*Table 1. Architectural rules considered by layering reconstruction approaches.*

|  | Abstraction rules | | Re­sponsi­bility rule | Transver­sality rule | Protection Against Varia­tions rules | |
|---|---|---|---|---|---|---|
|  | LAU rule | ILD rule |  |  | Interfacing rule | Stability rule |
| Sarkar et al. [15] | ✓ | ✓ | ✓ |  |  |  |
| Sarkar et al. [89] | ✓ | ✓ |  |  |  |  |
| Sangal et al. [28] | ✓ | ✓ | ✓ |  |  |  |
| Sangal and Waldman [102] | ✓ | ✓ | ✓ |  |  |  |
| Sangal et al. [40] | ✓ | ✓ | ✓ |  |  |  |
| Laval et al. [16] | ✓ | ✓ |  |  |  |  |
| Laval et al. [125] | ✓ | ✓ |  |  |  |  |
| Sora et al. [64] | ✓ | ✓ | ✓ | ✓ |  |  |
| Bischofberger et al. [41] | ✓ | ✓ | ✓ |  | ✓ |  |
| Bourquin and Keller [20] | ✓ | ✓ |  | ✓ |  |  |
| Kazman and Carrière [5] | ✓ | ✓ |  |  |  |  |
| Schmidt et al. [38] | ✓ | ✓ | ✓ |  |  |  |
| Hautus [19] | ✓ | ✓ |  |  |  |  |
| Pruijt and Wiersema [96] | ✓ | ✓ |  |  |  |  |
| El Boussaidi et al. [8] | ✓ | ✓ | ✓ | ✓ |  |  |
| Boaye Belle et al. [33] | ✓ | ✓ |  |  |  |  |
| Boaye Belle et al. [39] | ✓ | ✓ | ✓ |  |  |  |
| Boaye Belle et al. [18] | ✓ | ✓ |  |  |  |  |
| Boaye Belle et al. [37] | ✓ | ✓ | ✓ |  |  |  |
| Müller and Uhl [131] | ✓ | ✓ | ✓ | ✓ | ✓ |  |
| Müller et al. [47] | ✓ | ✓ | ✓ | ✓ | ✓ |  |
| Kienle and Müller [57] | ✓ | ✓ | ✓ | ✓ | ✓ |  |
| Andreopoulos et al. [66] | ✓ | ✓ |  |  |  |  |
| Constantinou et al. [98] | ✓ | ✓ |  |  |  |  |
| Cai et al. [115] | ✓ | ✓ | ✓ |  |  |  |
| Harris et al. [6] | ✓ | ✓ | ✓ |  |  |  |
| Stoermer et al. [3] | ✓ | ✓ | ✓ |  |  |  |
| Duenas et al. [121] | ✓ | ✓ | ✓ |  | ✓ |  |
| Chagas et al. [128] | ✓ | ✓ | ✓ |  |  |  |
| Marinescu [126] | ✓ | ✓ | ✓ |  |  |  |
| Riva et al. [130] | ✓ | ✓ |  |  |  |  |
| Riva [129] | ✓ | ✓ | ✓ | ✓ | ✓ |  |
| Mohammed and Fawcett | ✓ | ✓ | ✓ |  |  |  |
| Ossher [132] | ✓ | ✓ | ✓ |  |  |  |
| Van Deursen et al. [122] | ✓ | ✓ |  |  |  |  |
| Cole et al. [123] | ✓ | ✓ |  |  |  |  |
| Goldstein and Segall [137] | ✓ | ✓ | ✓ |  | ✓ |  |

*5.1.3. On the reconstruction of architectures exceptionally violating the layering rules*

Some of the surveyed approaches support the reconstruction of layered architectures that are tolerant to layering violations (e.g., [15, 18, 33, 37, 39, 89]) such as skip-calls, back-calls and/or cyclic dependency violations. Thus, the so-reconstructed architectures do not strictly comply with the layering rules they are leveraging during the reconstruction process. Such a tolerance is usually meant to reflect a conscious architectural decision made by designers [89] when creating a software system. For instance, some designers strictly enforce the abstraction rules—and more specifically the Incremental Layer Dependency rule—while others, to address concerns such as lack of performance and flexibility inherent in the layered pattern [21], tend to be lax on these rules, thus introducing violations such as skip-calls. This is consistent with a study by Bischofberger et al. [41] on fifty software systems which showed that it is hard to implement medium-size software systems without introducing architectural violations. This is also corroborated by our recent experiments on the recovery of systems such as JHotDraw [83], which is considered as an example of a well-designed layered framework.

To account for these violations, some architecture reconstruction approaches propose parameters to be tuned by designers to indicate the extent to which they enforced the layering rules at design-time (e.g. [18, 33, 37, 64, 66]). For example, user input could instruct a layering reconstruction technique to reconstruct an architecture where the back-calls are negligible compared to the other layer dependencies while when skip-calls are twice-more tolerated than intra-layer dependencies, this led to good layering results in [18, 37]. Along the same line, we assume that the extent to which the Protection Against Variations rules are enforced in a layered architecture may depend on the interfacing level that designer's privilege, i.e., a gray-box approach versus a black-box approach, or even a lack thereof (i.e., white-box approach). Hence, if the designers follow a gray-box approach, it could be handy to use that information as input to instruct the reconstruction technique to use, for instance, design patterns techniques to recover Facade pattern [59] instances implementing the interfaces of the layers' components.

Approaches such as [28] also support the specification of design rules reflecting the architectural constructions allowed in the analyzed system. The application of these rules enables the detection of architectural violations and ensures that during the evolution of a system, the conceptual architecture remains consistent with its concrete architecture.

### 5.2. Supporting the reconstruction process using the layered principles

The techniques used by existing software architecture reconstruction approaches can be classified into automatic, semi-automatic or quasi-automatic [4]. In the context of the layered architecture reconstruction, approaches are either semi-automatic (e.g., [15, 16, 28, 29, 102, 128]) or quasi-automatic (e.g., [38, 64, 98]). The semi-automatic approaches are usually based on abstraction techniques such as queries (e.g., [47]), on graph-theory techniques (e.g., [15, 29, 64, 121]) or on (meta-)heuristics (e.g., [8, 18, 19, 38]). Quasi-automatic approaches usually rely on clustering (e.g., [38, 64, 115]). Even though these techniques are derived from architectural rules, they do not straightforwardly leverage the targeted rules through an implementation as rule-based (expert) systems, which seems to be one of the most obvious ways to implement such approaches.

Academic and industrial tools have been developed to (semi-)automate the reconstruction of layered software architecture. These tools usually allow reconstructing layered software architectures, visualizing the reconstructed artifacts (e.g., layering metrics, reconstructed layered architectures, and architectural violations) and refining them. The most notable industrial tools are:

- Lattix [97, 160, 164]
- Structure101 [167]
- Ndepend [85]

Research tools include:

- Mohammed and Fawcett's tool [162]
- Ozone [16, 125]
- ReALEITY that supports the approaches discusses in [18, 33, 37]
- Rigi [57, 47].

In the following, we further describe how software architecture reconstruction approaches rely on each layering principle to guide the reconstruction.

*5.2.1. The abstraction principle in the software architecture reconstruction literature*

The abstraction principle has led to many algorithms that reconstruct layered architectures based on a depth-traversal of dependency graphs built from the studied system (e.g., [5, 16, 18, 29]). Existing approaches can be classified according to the heuristics they use:

- **Heuristics to handle cyclic dependencies**: In accordance with the Incremental Layer Dependency rule, several approaches focused their effort on proposing methods and heuristics to handle entities involved in cyclic dependencies (e.g., [15, 16, 28, 129]). The rationale is that one of the most common features in large software systems is the presence of cyclic dependencies [4, 84]. For example, in a system like ArgoUML, the largest cycle comprises almost half of the system's packages. In jEdit, the largest cycle contains almost two-thirds of the packages [84]. The presence of cyclic dependencies might hinder the reconstruction process, particularly for software architectures that are thought to comply with the layered pattern. Indeed, we can no longer assess abstraction levels. Thus, many approaches start by getting rid of these cycles, before applying the layering rules. Roughly speaking, approaches that handle such cycles use some heuristic to build a directed acyclic graph from the elements of the analyzed system. The acyclic graph is then used to build the layering of the system. Depending on the heuristic used, the resulting layering may vary.

- o *Heuristics based on highly connected components*: such approaches (e.g., [15, 28]) usually consist in detecting strongly connected components and then assigning them to layers based on the usage flow. They usually generate few layers because they implicitly favor the grouping of similar responsibilities exhibited by components involved in the same cycle.
- o *Heuristics based on cycle breaking*: to resolve the cyclic dependency problem, such approaches detect and remove undesired dependencies causing cyclic dependencies before assigning modules to layers (e.g., [16, 19, 125]). Such approaches usually generate several layers because they promote adjacent dependencies between the layers.
- **Heuristic tuning**: to obtain more accurate results, some layering recovery approaches usually rely on parameters that allow tuning heuristics based on the layering pattern rules and on how strictly the architect applied them when designing her system. For instance, based on the incremental dependency rule, a recent work [18, 33, 37] derived a set of layer dependency attributes and constraints. These helps translate the reconstruction of layered architectures into a quadratic assignment problem (QAP) [18, 33, 37], a well-established combinatorial optimization formulation which has been used to model problems such as layout design or resource allocation. They solved that QAP using search-based algorithms. Their formalization of the reconstruction as an optimization problem represents a more general way of expressing and tuning the different heuristics used by other work.
- **Heuristics based on fan-in and fan-out dependencies:** with such approaches (e.g., [38]), a module that does not have fan-out dependencies is assigned to the lowest-level layer and conversely a module that does not have fan-in dependencies is assigned to the highest-level layer. Modules that are used by many other classes are grouped in the lowest layer while modules that rely on many other modules are grouped in the highest layer. The remaining modules are usually assigned to a middle layer.

*5.2.2. The responsibility principle in the software architecture reconstruction literature*

The responsibility principle is related to modularity which has already been subject to many studies (e.g., [8, 9, 11]). Most work on architecture reconstruction using clustering techniques focuses on:

- **Coupling minimization:** Some approaches (e.g., [9, 11, 8]) aim at decomposing systems into components while minimizing the coupling between resulting components and maximizing the cohesion of each component based on the structural dependencies between entities.
- **Lexical similarities:** These use lexical similarities between elements of the system to identify components or modules in systems (e.g., [15, 33, 45, 46, 47, 48, 115]). Most such approaches, focusing on the layered pattern, rely on lexical analysis of the source code. They use elements such as identifiers and comments to compute lexical similarities between source code entities. Approaches such as [15, 37, 47, 57] relied on lexical similarities as a complement to the structural information retrieved from the systems at hand. The lexical information used as input by some of these approaches is obtained by applying Information Retrieval techniques such as LDA (Latent Dirichlet Allocation) [91] on the identifiers and comments comprised in the source code.
- **Strongly connected components identification:** This is related to the cyclic dependency principle. Some approaches (e.g., [15, 28, 137, 162]) rely on techniques to detect strongly connected components to identify modules that are cyclically involved. The modules are usually assigned to the same layer during the reconstruction process. This usually implies that such modules share a common layer responsibility.

*5.2.3. The transversality principle in the software architecture reconstruction literature*

Exploiting the Transversality principle when reconstructing layered architectures eases reconstruction by creating a transverse component comprising omnipresent modules which might then be excluded from the reconstruction process. Doing so is a practice advocated and applied by many approaches (e.g., [47, 8]). The main argument against omnipresent components is that they tend to obscure the reconstructed architecture [47, 54]. Hence, grouping them eases the reconstruction [47].

Some approaches support the identification of omnipresent modules while reconstructing layered architectures (e.g., [8, 47]). However, although some transversality is recommended when designing layered systems, existing reconstruction approaches focusing on the layered pattern barely, implicitly or explicitly, rely on that rule while reconstructing. Work that does so might include [8] and [64] through the possibility of combining a layering reconstruction technique with clustering techniques supporting the identification of omnipresent modules. An exception to that trend is also found in the Rigi tool [47, 57] that explicitly supports the exclusion of omnipresent modules from the layering process.

*5.2.4. The protection against variations principle in the software architecture reconstruction literature*

We can think of work using dominance trees (e.g., [55, 62]) as general reconstruction techniques that implicitly use the Interfacing rule, i.e., the identification of an entry point leads to the identification of a whole subsystem. But when it comes to the layered pattern, although various papers advocate enforcing the Protection Against Variations rules when designing layered architectures, it is not common practice to rely on these rules when reconstructing architectures. Rigi ([47, 57]) diverges from that trend by reconstructing interfaces among the reconstructed subsystems, thus relying on the Interfacing rule. Interestingly, the Protection Against Variations rules guided the framework presented in [42], which analyzes layered architectures to see if layers restrict access to their services through interfaces. Results of the study in [42] suggest that developers tend to define and use input interfaces but not output interfaces.

To detect violations of the Factory pattern in an abstract graph representing a system, Goldstein and Segall [137] identified classes that violate the usage of factory interfaces by directly accessing the implementation that these interfaces use. This detection approach can be combined with the reconstruction of the layered architectures. It relates to some extent to the Interfacing rule.

## 6. COMPARISON WITH RELATED WORK

Several authors have surveyed the layered pattern fundamentals, software architecture reconstruction approaches, or a related theme. In the following, we discuss their work, and outline their limitations.

### 6.1. Related surveys

In 2009, Savolainen and Myllarniemi [70] carried out a systematic literature review to survey the literature found in five databases about software architecture layers. Their survey focused on eleven primary studies as well as a couple of selected books. It showed that the usage of the layered architectural pattern varies greatly based on the targeted contexts. Some papers discuss the layers in the context of distributed communication systems, while others discuss layers in the context of object-oriented systems. Their survey also showed that, despite being a well-established pattern, the research carried on software architectures compliant with the layered pattern is very scarce. Several papers discuss layering in some form or another, but there are just a few of them that address the foundation of layered architectures. Most of the books on layered architectures are either based on industrial practice or on the same seminal work discussed in [71] and [72]. Their survey also showed that some papers are vague when it comes to the description of the principles used to decompose and group elements into layers. Based on the surveyed literature, the survey acknowledges that layering should support separation of concerns as well as reuse and be guided by the abstraction principle. But the authors did not explain what these concepts convey.

In 2010, Ducasse and Pollet [4] surveyed 181 software architecture reconstruction (SAR) approaches based on a five-criteria taxonomy. The five criteria covered by their taxonomy are the following: the inputs, the goals, the processes, the outputs and the techniques characterizing SAR approaches. In particular, the inputs refer to the parameters of the architectural reconstruction process. Inputs include non-architectural data, architectural data such as architectural patterns or a combination of both types of data. The non-architectural data include structural, textual, and dynamic information as well as human expertise. The techniques refer to the degree of automation of the architectural reconstruction process. A technique can be quasi-manual, semi-automatic or quasi-automatic. With a quasi-manual technique, the user manually identifies architectural elements using a tool that helps understand the reconstructed abstractions. A semi-automatic technique automates the repetitive steps of the reconstruction but requires that a user manually instructs the tool to iteratively refine architectural elements to reconstruct abstractions. With a quasi-automatic technique, the user guides the iterative reconstruction of abstractions, but the tool controls it.

In 2013, through their TSLR (Typology of Software Layer Responsibility), Pruijt et al. [105] proposed an inventory of the common responsibility types that can be found in the software of business information systems complying with the layered pattern. Their classification scheme leverages leading literature about software layers and revolves around three levels of abstraction varying in accordance with the granularity of the responsibilities. Such responsibilities include the Consumer Interface Responsibility, the Task Specific responsibility, and the Domain Generic responsibility. To support the application of the TSLR in practice, they also proposed an RTT (Responsibility Trace Table) that maps the TSLR-responsibilities to the application-specific layers (e.g., presentation layer, domain layer). Pruijt et al. also acknowledged that the number of layers as well as their respective names may vary from a system to another, that the responsibilities per layer may vary, and that a responsibility can be a composite, i.e., comprise several sub-responsibilities. This resulted in a TSLR meta-model matching the structure of the TSLR.

The literature often uses the expressions "architectural styles" and "architectural patterns" interchangeably [27, 44], even though there are some differences in their formalization of software architectures constructs [27]. Harrison et al. [86] pointed out that architectural patterns support the efficient documentation of some of the most important design decisions. Along the same lines, Tofan et al. [73] published a systematic mapping study aiming at investigating the state of research on software architectural decisions. That study included 144 papers published between January 2002 and January 2012. They classified architectural decisions into three classes: existence decisions, property decisions and executive decisions. Existence decisions indicate the existence of some artifacts (e.g., the system will consist of three layers). Property decisions relate to factors (e.g., design rules, design guidelines, as well design constraints) that influence several elements of a software system. Executive decisions relate to the choice of processes, tools, or technologies. The authors also identified possible future research topics. These include domain-specific architectural decisions (e.g., mobile), realization of specific quality attributes (e.g., reliability or scalability), uncertainty in decision-making, as well as group decisions.

In 2018, Muccini and Moghaddam [155] carried out a systematic mapping study involving 63 papers. Their objective was to harmonize and integrate the literature on both IoT systems and software architecture patterns through the definition, identification, categorization, and re-design of various architectural patterns and patterns that can ease the implementation of IoT architectural characteristics. Their study showed that primary studies used one or several overlaid patterns to design their IoT architecture. These IoT architectural patterns include: the layered pattern, the cloud-based pattern, the service-oriented pattern, and microservices. Among these patterns, the layered architectural pattern is the most used.

The same year, Pahl et al. [44] claimed that there is a need for a specific software architectural pattern for cloud-based software systems to support the development and operation of continuous service systems. In this regard, based on a survey of the relevant literature, they defined a cloud architectural pattern with various principles (e.g., service-orientation, virtualization, adaptation, and uncertainty) and patterns (microservices, quality models at runtime, control loop, and controller architecture) it conveys. In their work, they considered service-orientation as a meta-principle (i.e., grouping of principles) that: (1) defines the SOA architectural pattern, and (2) relies on loose coupling, modularity, and layering as guiding principles. As such, software layers (usually the presentation, business logic, and data management layers) structure the code from logical and development points of view, while physical tiers correspond to the physical places where an application runs. In the cloud, a layered system is usually mapped onto physical tiers and their native services. Layers are therefore deployed and executed on *tiers*. These tiers are realized by Infrastructure-as-a-service (IaaS), platform-as-a-service (PaaS) or software-as-a-service (SaaS).

**6.2. Gaps in the literature**

As in [44], [105], and [155], one the goals of our systematic literature reviews is to harmonize and merge the literature on one or several architectural patterns to ease their use. As in [44], we have addressed both commonly accepted and recent concerns regarding the surveyed theme such as the poor definition of the layered pattern and the lack of uniformity in its descriptions (e.g., [70, 105]). However, existing surveys (i.e. [70, 105]) covering the layered pattern usually either focus on the description of a single principle (e.g. [105]) or acknowledge the existence of a subset of layering principles without describing the essence of such principles (e.g. [70]). For instance, Pruijt et al. [105] focused on a single layering principle: the responsibility principle.

We distinguish ourselves from the work discussed above in that we survey more extensively the literature on the layered pattern, we cover more principles and rely on more criteria to survey that literature: the essence of the layering principles, their illustration and their use in the literature focusing on software architecture reconstruction. This allows us to propose: (1) a synthesis of rules embodying the essence of layering principles; (2) a set of diagrams that can be used to depict layered software architectures based on the principle at hand; and (3) an investigation of the usage of layered style rules in the software architecture reconstruction.

## 7. LIMITATIONS AND THREATS TO VALIDITY

As Figure 2 indicates, some of the principles we discuss are not systematically orthogonal i.e., independent from each other. For instance, the protection against variations rules are closely related to the responsibility rule. Readers may have expected us to present a more formalized version of the rules in a way that make them as independent of each other as possible, analogous to axioms in formal mathematical approaches. Thus, not formalizing such principles in our work may be perceived as a limitation. Still, our work does not focus on the formalization of such principles, but on their exploration in the literature and their synthesis. Some other work did focus on such a formalization: this includes [33], and [37].

As stated above, the literature on the layered patterns interprets its principles using different interpretations of the layering paradigm, that usually lack uniformity. The resulting architecture descriptions may often clash. We have tried to make existing principles more uniform by aggregating them as a set of general design rules that they may then apply indiscriminately when analyzing layered software architectures. Still, a rule violation in one interpretation of the layered pattern may be considered as valid in some other interpretations. For instance, "intra-dependencies" are accepted in some interpretations of the layered pattern, whereas in other interpretations, they are considered as undesirable. Still, we have made our best to aggregate these rules in a way that reflect their degree of acceptance and/or intolerance depending on the considered interpretation.

The manual search and database-driven strategies are the most used search strategies to identify the research [170]. We have therefore used them to identify the search in our two SLRs. Relying solely on these two search strategies may have undermined the completeness of our reviews, since we may have missed a couple of papers that could have been found by relying on additional search strategies. Such additional search strategies include snowballing [50], which is a search strategy that is increasingly used by the literature. We did not rely on additional search strategies because we tried to find a good balance between the number of primary studies to include in our reviews and the time required to implement different search strategies. Relying on the chosen search strategies to identify the search therefore seems like a good compromise, especially since it yielded a very high number of primary studies. Besides, we have mitigated the completeness threat by searching primary studies in databases that are very popular and well-known, and therefore comprise an extensive variety of scientific articles. Scopus for instance, is a quite complete database, especially since it also indexes publications from several well-known publishers such as Elsevier and Springer and may therefore comprise several leading publications [51]. Noteworthy, authors such as Kitchenham [51], only relied on Scopus to perform their mapping studies.

In addition, the choice to rely on these two search strategies to identify the search is in accordance with the literature on systematic literature reviews and systematic mapping studies (e.g., 170]). Systematic literature reviews and systematic mapping studies share some common features (e.g., searching and study selection). Some of their common guidelines can therefore be useful to carry out systematic literature studies. When carrying out their systematic mapping study on systematic mapping studies, Petersen et al. [170] found out that the most used search strategy to identify the search (i.e., primary studies) is the database-driven search. They also concluded that manual searches are beneficial, and even more effective to identify the search. Besides, some mapping studies they investigated only relied on the manual search as their unique approach to identify the search. Finally, Petersen et al. stressed out that, relying on multiple search strategies to identify the search can be quite time-intensive: a balance between timely availability of the information and achieving a good overview of the research area therefore needed to be taken in consideration when identifying the search.

## 8. CONCLUSION AND PERSPECTIVES

### 8.1. Discussion and perspectives on the layered architectural pattern

In this paper, we have systematically surveyed the literature describing the layered pattern and synthesized the architectural rules characterizing layers in software applications that follow the layered pattern, as well as approaches to rearchitecting systems following that pattern.

The layered pattern has stood the test of time. However, its principles have barely evolved, are loosely defined, vary from primary studies to primary studies, and lack uniformity.

We explained in Section 1 how early architecture reconstruction work relied on general modular design principles, such as high cohesion and low coupling, to identify the components of an architecture (e.g. [9, 11, 8]).

The rules we discuss in this paper are specific to the layered architectural pattern and are consistent with more recent architecture reconstruction work (e.g. [15, 33]). While most researchers agree on what a layered architecture is meant to achieve, different authors have presented subtly different rules to uncover layers as reported in Table 1 (see Section 5). This reflects different ways of attaining the objectives of the pattern. In this context, the synthesized rules are not simply a collection of rules gleaned from the literature but identifies four principles that cover these rules: (1) the abstraction principle, (2) the responsibility principle, (3) the transversality principle, and (4) the protection against variation principle.

We presented (in Section 4) a review of the rules derived from these principles. And since the reconstructed architecture allows carrying out diverse analyses for purposes such as the identification of architectural technical debt, we then presented a second review (in Section 5) existing software architectural approaches in the light of the rules synthesized in the first review.

This second review reflects the complexity of uncovering layers within a layered legacy system, which is due to several factors:

First, as an *architectural pattern*, the layered pattern is loosely defined, more so than other patterns such as the virtual machine pattern or the publish and subscribe pattern. For instance, in terms of architectural pattern 'theory' (see Section 2.1), the layered pattern has a single 'component type': the software layer (i.e., in layered systems, the components are layers). Strictly speaking, a layer is not a *component* in the usual sense of a component: it exhibits neither the functional cohesion nor the 'runtime' unity/cohesion of a typical component [23]. As explained in section 2, a layer will typically include several relatively independent tasks that may only share the *abstraction level*, as opposed to functional cohesion. In terms of runtime behavior, some function executions may cross many layers, and involve one or several 'modules' from each layer. For example, the execution of most web services will minimally integrate code from the application layer, the business layer, and the data layer.

Second, the notion of a layer is mostly an 'abstraction of the mind'. There are no language constructs that support the concept of a layer [5], unlike the concept of a *service*, for example, where we have different IDLs (Interface Definition Languages), depending on the target technology (e.g., *WSDL*, for *web services*, *REST* APIs, for *REST* services, etc.). The often-used example of the extremely well-behaved OSI stack, whose "good behavior" was made possibly thanks to: (1) a narrow functional scope, in particular for the first five layers, and (2) stringent, do-or-die interoperability constraints that are rather the exception than the rule.

Accordingly, the rules that we discuss in this paper may include a subjective user's input (i.e., an indication of the degree of tolerance of some rules), which helps the designer select the 'degree of conformance' relative to the ideal (OSI like) scenario. On a more conceptual level – giving more credence to the lack of consensus on the definition of the layered pattern – we think that intra-level dependencies should neither be forbidden nor discouraged. And consequently, the abstraction level should not be taken as the number of calls down to the physical level. We don't have a better/more "operational level" of "abstraction", but if we think of the ideal/idealized layering pattern, as exemplified by the (ideal) OSI hierarchy, we can think of a layered application as a hierarchy of "virtual machines" each one offering a "language" to the levels above it, and "relying on a lower-level virtual machine". This is in accordance with the literature (e.g., [2]) and this may make it easier to think of an "abstraction level" as a "virtual machine" that should have all the "commands" (interface) needed to execute "programs" above it (level j-1). This view may lead to a more "semantic understanding" of LAU rule, ILD rule, and responsibility principle/rule. Also, "skip-calls" become "symptoms" of a "command" at abstraction level $j+1$ that is exposed as is at abstraction level $j$, without wrapping it (or, with a simple call forwarding, which would "hide" the skip-call).

As future work, we aim at synthesizing the rules conveyed by other architectural patterns such as the blackboard and the pipe-and-filter patterns. As the structural violations infringing the architectural rules are treated as architectural technical debt (ATD) – a dimension of technical debt [67, 108, 147, 164, 168] – future work will also be dedicated to the creation of a catalogue of architectural violations on which one can rely to derive metrics to assess ATD in existing software systems. Such metrics will help assess the impact of such violations on software architectures. This aspect is particularly important since ATD is challenging to uncover and, when not properly managed, can lead to very expensive repercussions on the system being developed and maintained [99, 120].

### 8.2. Software architecture reconstruction in general: some additional perspectives

Our analysis of layered architecture reconstruction approaches (see Sections 5) and of several software architecture reconstruction approaches in general (e.g., [9, 10, 19, 28, 52, 53, 54, 55, 56, 70]) drove us to propose some related perspectives to consider. We discuss them to conclude this paper.

#### 8.2.1. Need for language-and-platform independent reconstruction techniques

Most of the architecture reconstruction approaches (e.g., [19, 28, 47, 54, 56]) are language and platform-dependent and do not use a standard representation of the data of the system under analysis [8]. Therefore, the resulting tools are not able to interoperate with each other [80]. To tackle this issue, architecture reconstruction approaches should systematically rely on a standard representation of the data used as input to the reconstruction process. This perspective is notably supported by the OMG that has recently undertaken an effort to standardize software reengineering through the Architecture-Driven Modernization taskforce [83, 107]. In this context, a metamodel such as the KDM (Knowledge Discovery Metamodel) standard [78] can be very helpful since it enables to represent legacy systems' artifacts in a language and platform-independent manner.

#### 8.2.2. Need for reconstruction approaches for large heterogenous systems

The next generation of software architecture reconstruction techniques needs to be more interdisciplinary, i.e., to combine various techniques from different disciplines to be able to deal with the reconstruction of large software-intensive systems. As stated by Arias and Avgeriou [78], such systems usually consist of millions of lines of code developed using many different programming languages, which results in a heterogenous implementation. These systems are usually exposed to several changes throughout the years, which increases the likelihood that they deviate from their original design thus causing a severe mismatch with their concrete architecture. Besides, the legacy components that these systems comprise are the result of large investments and embed multidisciplinary knowledge spread among their development organization's experts. Hence, combining traditional software engineering (SE) based reconstruction techniques with artificial Intelligence based techniques using machine learning (as in [98]), Expert systems or even recommender systems (among others) may foster the development of semi-automatic reconstruction that will efficiently leverage and (semi-)automate the knowledge from the experts when reconstructing software-intensive system architectures.

This perspective is supported by a very recent study [75] showing that machine-learning based techniques can be effectively used to help reconstruct software architectures. Note that combining SE techniques with expert systems techniques to support architecture reconstruction requires finding suitable data structures to properly represent expert knowledge on architecture reconstruction and to capture the expert's reasoning capabilities through adequate inference rules. Transposing work such as [76] to the reconstruction of software architectures could help find good insights on the matter.

#### 8.2.3. Need for comparison baselines

The next generation of approaches supporting software architecture reconstruction should be able to rely on a vast repository hosting various ranges of ground-truth architectures when validating their work. Even though a couple of works address the reconstruction of ground-truth architectures (e.g., [14, 84]), there is still a lack of comparison baselines in the software architecture community [29, 82]. This is largely because obtaining the ground-truth architectures is arduous and time-consuming. Further, the idealized architecture that architects and engineers have in mind often deviates from the concrete one [14]. In this context, creating ground-truth architectures would help populate a repository housing the architectures used in experiments carried out by the research community. This would give an opportunity to those who access the repository to comment its content and to possibly propose some adjustments to improve it [29]. Note that obtaining the ground-truth of architectures also calls for the establishment of a sound protocol that will allow involved architects to reconcile their respective proposed architectures to obtain the final decomposition of the system at hand in a timely manner.

#### 8.2.4. Toward an ideal pattern-specific reconstruction process

The ideal reconstruction process guided by an architectural pattern may be the one that leverages all (or most of) the principles it conveys. For instance, in the context of the layered pattern, we think that the ideal software architecture reconstruction process may be the one that will take as input the analyzed system's information as well as the user's input, and then rely on the pattern rules to reconstruct the system's layered architecture.

That process will first need to apply the transversality rule to identify subsystems or modules that contain omnipresent functions. These modules can be put in a transverse component (i.e., sidecar).

The responsibility and protection against variations rules can then be applied separately or combined on the system from which the omnipresent subsystems are excluded (e.g., by ignoring their dependencies with the other modules). The protection against variations rules are closely related to the responsibility rule. The responsibility rule will help identify cohesive components/modules while the interfacing rule will help identify which of these components share the same interfaces. Once the responsibilities are identified and their shared interfaces are defined, the reconstruction process can then apply the abstraction rules to lay out the responsibilities through the layers. In the context of the layering reconstruction, it is mandatory to first identify the components of the system by applying the responsibility rule and then apply the abstraction rules to assign each component to the appropriate layer.

Once the layers are reconstructed, the user or domain experts can visualize them and refine them based on their knowledge of the system at hand. This can improve the quality of the resulting layering by adding information that is absent in the code [8].

Our analysis of existing work (see Section 5) showed that very little work has combined most of these rules either because of the relaxation of the layered pattern essence throughout the years or the complexity associated with the implementation of some of these rules as part of the reconstruction process. This calls for the investigation of more advanced techniques to effectively translate architectural rules into a reconstruction process.

## APPENDIX

*Table 2. Primary studies included in both SLRs*

| Identifier | Reference No | Year | Authors | SLR |
|---|---|---|---|---|
| 1. | [2] | 1993 | Garlan and Shaw | First SLR |
| 2. | [8] | 2012 | El Boussaidi et al. | Both SLRs |
| 3. | [13] | 2005 | Larman | First SLR |
| 4. | [15] | 2009 | Sarkar et al. | Both SLRs |
| 5. | [16] | 2013 | Laval et al. | Both SLRs |
| 6. | [18] | 2015 | Belle et al. | Both SLRs |
| 7. | [19] | 2002 | Hautus | Both SLRs |
| 8. | [20] | 2007 | Bourquin and Keller | Both SLRs |
| 9. | [21] | 1996 | Buschmann et al. | First SLR |
| 10. | [22] | 2002 | Clements et al. | First SLR |
| 11. | [24] | 2002 | Eeles | First SLR |
| 12. | [25] | 2000 | Bachmann et al. | First SLR |
| 13. | [26] | 2007 | Bachmann et al. | First SLR |
| 14. | [27] | 2005 | Avgeriou and Zdun | First SLR |
| 15. | [28] | 2005 | Sangal et al. | Both SLRs |
| 16. | [30] | 1980 | Zimmermann | First SLR |
| 17. | [31] | 2004 | Paris | First SLR |
| 18. | [33] | 2013 | Belle et al. | Both SLRs |
| 19. | [37] | 2016 | Belle et al. | Both SLRs |
| 20. | [39] | 2014 | Belle et al. | Both SLRs |
| 21. | [40] | 2005 | Sangal et al. | Both SLRs |
| 22. | [41] | 2004 | Bischofberger et al. | Both SLRs |
| 23. | [42] | 1998 | Laguë et al. | First SLR |
| 24. | [43] | 2013 | Pruijt et al. | First SLR |
| 25. | [47] | 1993 | Müller et al. | Both SLR |
| 26. | [57] | 2010 | Kienle and Müller | Both SLRs |
| 27. | [60] | 2003 | Buschmann and Henney | First SLR |
| 28. | [63] | 1995 | Kruchten | First SLR |
| 29. | [64] | 2010 | Șora et al. | Both SLRs |
| 30. | [71] | 1968 | Dijkstra | First SLR |
| 31. | [72] | 1972 | Parnas | First SLR |
| 32. | [74] | 2010 | Clements et al. | First SLR |
| 33. | [86] | 2007 | Harrison et al. | First SLR |
| 34. | [89] | 2006 | Sarkar et al. | Both SLRs |
| 35. | [94] | 2009 | Wong et al. | First SLR |
| 36. | [95] | 1979 | Parnas | First SLR |
| 37. | [96] | 2017 | Pruijt and Wiersema | Both SLRs |
| 38. | [98] | 2011 | Constantinou et al. | Both SLRs |
| 39. | [100] | 2017 | Bouihi and Bahaj | First SLR |
| 40. | [101] | 2006 | Zewdie and Carlson | First SLR |
| 41. | [102] | 2005 | Sangal and Waldman | Both SLRs |
| 42. | [103] | 2008 | Connolly | First SLR |
| 43. | [104] | 2001 | Cunningham and Wang | First SLR |
| 44. | [105] | 2013 | Pruijt et al. | First SLR |
| 45. | [106] | 1995 | Hayes-Roth et al. | First SLR |
| 46. | [107] | 2017 | Landi et al. | First SLR |
| 47. | [109] | 2000 | Nord | First SLR |
| 48. | [110] | 2000 | Spicer | First SLR |
| 49. | [111] | 2015 | Mariani et al. | First SLR |
| 50. | [113] | 2014 | Pruijt and Brinkkemper | First SLR |
| 51. | [114] | 2016 | Pruijt et al. | First SLR |
| 52. | [115] | 2009 | Cai et al. | Both SLRs |
| 53. | [116] | 2017 | Muram et al. | First SLR |
| 54. | [117] | 2010 | Sobernig and Zdun | First SLR |
| 55. | [118] | 2014 | Samrongsap and Vatanawood | First SLR |
| 56. | [121] | 1998 | Duenas et al. | Both SLRs |
| 57. | [123] | 2005 | Cole et al. | Both SLRs |
| 58. | [124] | 2018 | Konara et al. | Both SLRs |
| 59. | [125] | 2010 | Laval et al. | Both SLRs |
| 60. | [126] | 2006 | Marinescu | Both SLRs |
| 61. | [127] | 2005 | Zdun and Avgeriou | First SLR |
| 62. | [128] | 2016 | Chagas et al. | Both SLRs |
| 63. | [129] | 2000 | Riva | Both SLRs |
| 64. | [130] | 2004 | Riva et al. | Both SLRs |
| 65. | [131] | 1990 | Muller and Uhl | Both SLRs |
| 66. | [133] | 1985 | Rajlich | First SLR |
| 67. | [134] | 1992 | Batory and O'malley, | First SLR |
| 68. | [135] | 2007 | Concas et al. | First SLR |
| 69. | [136] | 2009 | Deiters et al. | First SLR |
| 70. | [137] | 2015 | Goldstein and Segall | Both SLRs |
| 71. | [138] | 2013 | Herold and Rausch | First SLR |
| 72. | [139] | 2008 | Kamal et al. | First SLR |
| 73. | [140] | 2009 | Lockemann and Nimis | First SLR |
| 74. | [141] | 2007 | Melton and Tempero | First SLR |
| 75. | [142] | 2008 | Zdun and Avgeriou | First SLR |
| 76. | [144] | 2005 | Sarkar et al. | First SLR |
| 77. | [145] | 2007 | Sarkar et al. | First SLR |
| 78. | [151] | 2000 | Garlan | First SLR |
| 79. | [152] | 2010 | Saraiva et al. | First SLR |
| 80. | [153] | 2011 | Saraiva et al. | First SLR |
| 81. | [156] | 2017 | Kouamou and Kungne | First SLR |
| 82. | [162] | 2017 | Mohammed and Fawcett | Both SLRs |
| 83. | [171] | 2021 | Romhányi et al. | First SLR |
| 84. | [3] | 2003 | Stoermer et al. | Second SLR |
| 85. | [5] | 1999 | Kazman and Carriere | Second SLR |
| 86. | [6] | 1995 | Harris et al. | Second SLR |

| 87. | [38] | 2011 | Schmidt et al. | Second SLR |
| 88. | [66] | 2007 | Andreopoulos et al. | Second SLR |
| 89. | [122] | 2004 | Van Deursen et al. | Second SLR |
| 90. | [132] | 1984 | Ossher | Second SLR |